\documentclass[twocolumn,showpacs,preprintnumbers,amsmath,amssymb,aps]{revtex4}

\usepackage{graphicx}
\usepackage{dcolumn}
\usepackage{bm}
\newcommand\be{\begin{equation}}
\newcommand\ee{\end{equation}}

\begin{document}

\title{Lattice calculation of non-Gaussianity from preheating}

\author{Alex Chambers}
 \email{alex.chambers@imperial.ac.uk}
 \author{Arttu Rajantie}%
 \email{a.rajantie@imperial.ac.uk}
\affiliation{%
Department of Physics,
Imperial College London,
Prince Consort Road,
London SW7 2AZ,
UK}%

\date{22 October 2007}

\begin{abstract}
If light scalar fields are present at the end of inflation, their
non-equilibrium dynamics such as parametric resonance or a phase
transition can produce non-Gaussian density perturbations. We show how
these perturbations can be calculated using non-linear lattice field
theory simulations and the separate universe approximation. 
In the massless preheating model, we find that some parameter values are excluded while others lead to
acceptable but observable levels of non-Gaussianity. This shows that preheating can be an important
factor in assessing the viability of inflationary models.

\end{abstract}

\pacs{98.80.Cq, 11.15.Kc}
\preprint{Imperial/TP/07/AC/01}

\maketitle


Predicting the observed nearly scale-invariant and Gaussian primordial density perturbations has been the greatest success of the slow-roll inflationary paradigm. However, these predictions are very generic and shared by a vast number of different inflationary models. Even the small observed deviation from scale invariance can be expressed purely in terms of slow roll parameters, and is therefore not enough to distinguish between models.

In contrast, if deviation from Gaussianity is observed, it will at least in principle provide much more information and could allow us to single out the correct model. The level of non-Gaussianity is commonly characterised using the parameter $f_{\rm NL}$, which was originally defined~\cite{Komatsu:2000vy} for a specific ``local'' type of non-Gaussianity by
\begin{equation}
\label{equ:fNLorig}
\zeta=\zeta_0-\frac{3}{5}f_{\rm NL}\left(\zeta_0^2-\langle\zeta_0^2\rangle\right),
\end{equation}
where $\zeta$ is the curvature perturbation and $\zeta_0$ is a Gaussian random field.
The definition of $f_{\rm NL}$ has since been extended to cover arbitrary non-Gaussian fields~\cite{Maldacena:2002vr,Boubekeur:2005fj}, in which case it also becomes scale-dependent. Even this generalised $f_{\rm NL}$ does not give a full description of the non-Gaussianity, but it is nevertheless a useful quantity.

Observational constraints on $f_{\rm NL}$ are currently not particularly strong~\cite{Creminelli:2006rz}, $|f_{\rm NL}|\lesssim 100$, but they are likely to improve significantly with the Planck satellite. Unfortunately, the level of non-Gaussianity produced during inflation in single-field models is very small~\cite{Maldacena:2002vr},
$|f_{\rm NL}|\ll 1$, and is unlikely to be ever observed.

In many inflationary models the end of inflation involves non-linear non-equilibrium dynamics, such as a phase transition or parametric resonance. It is natural to ask whether these processes can produce density perturbations and how non-Gaussian this contribution would be (see recent review \cite{Bassett:2005xm} and references therein). It is clear that causality prohibits generation of super-Hubble perturbations from nothing. However, if there is a light scalar field, it would acquire similar scale invariant perturbations during inflation to the inflaton field, and these may well be converted into density perturbations at the end of inflation.

In this letter, we show how the perturbations produced by non-equilibrium dynamics at the end of inflation can be calculated in a non-linear way using lattice field theory simulations. The method is very general and can be applied to any inflationary model in which the dynamics are dominated by bosonic fields. For concreteness, we will focus on a particular inflationary model known as massless preheating~\cite{Prokopec:1996rr,Greene:1997fu,Bassett:1999cg}. 
Recent approximate calculations~\cite{Enqvist:2004ey,Jokinen:2005by} have suggested that preheating could produce very large levels of non-Gaussianity in this model, and we show that this is indeed the case for large parts of the parameter space. We will also discuss at a general level the conditions required for effective generation of perturbations at the end of inflation.

The model consists of an inflaton field $\phi$ coupled to a massless scalar field $\chi$, with the potential
\begin{equation}
V(\phi,\chi)=\frac{1}{4}\lambda\phi^4+\frac{1}{2}g^2\phi^2\chi^2.
\end{equation}
During inflation, $\chi$ is approximately zero, and the model behaves the same way as the ordinary single-field $\phi^4$ chaotic inflation model. If $g^2/\lambda \lesssim O(1)$, the mass $m_\chi=g\phi$ of the $\chi$ field is less than the Hubble rate, and therefore the $\chi$ field acquires a similar scale-invariant spectrum of perturbations to the inflaton field $\phi$.

When inflation ends, the inflaton $\phi$ starts to oscillate with a decreasing amplitude. In terms of rescaled field $\tilde\phi=a\phi$ and rescaled conformal time $\tau$ defined by $d\tau=a^{-1}\lambda^{1/2}\tilde\phi_{\rm ini}dt$, the oscillations are approximately described by the Jacobi cosine function \cite{htbook},
$\tilde\phi(\tau)=\tilde\phi_{\rm ini}{\rm cn}(\tau,1/\sqrt{2}).$
This gives rise to an oscillatory mass term for the $\chi$ field. At linear level, a Fourier mode of the rescaled field $\tilde\chi=a\chi$ with wave number $k$ satisfies the Lam\'e equation,
\begin{equation}
\label{equ:lame}
\tilde\chi_k''+\left[
			\kappa^2+\frac{g^2}{\lambda}{\rm cn}^2(\tau,1/\sqrt{2})\right]
			\tilde\chi_k=0, 
\quad
\kappa^2=\frac{k^2}{\lambda\tilde\phi_{\rm ini}^2}.
\end{equation}

In the space of the two constant parameters $\kappa^2$ and $g^2/\lambda$, Eq.~(\ref{equ:lame}) has resonance bands in which the solution grows exponentially, $\tilde\chi_k(\tau)=e^{\mu\tau}f(\tau)$, where $\mu$ is known as the Floquet index and $f(\tau)$ is a periodic function (see Fig.~4 in Ref.~\cite{Greene:1997fu}). 
This means that energy is transferred rapidly from the inflaton $\phi$ to the $\chi$ field, which is known as preheating~\cite{Traschen:1990sw,Kofman:1994rk}.
The resonance bands stretch through the $(g^2/\lambda,\kappa^2)$ plane diagonally so that for any value of $g^2/\lambda$ there are some resonant modes, but when $g^2\ll\lambda$, the bands are narrow and the resonance is weak. Effective preheating therefore requires $g^2/\lambda\gtrsim 1$.

This model has been studied extensively using lattice field theory simulations~\cite{Prokopec:1996rr,Felder:2007ef}, but the focus has always been on field dynamics rather than on its effect on curvature perturbations. In a recent work~\cite{BasteroGil:2007mm}, gravitational effects were introduced at linearised level by coupling the fields to metric perturbations. This way, one can study their back-reaction to the field dynamics but, because only linear small-scale perturbations are included, not observable length scales or non-Gaussianity.

Our approach combines lattice field theory with the widely used separate universe approximation~\cite{Salopek:1990jq}. It states that points in space separated by more than a Hubble distance cannot interact and will therefore evolve independently of each other. As long as each Hubble volume is approximately isotropic and homogeneous, one can approximate them by separate Friedmann-Robertson-Walker (FRW) universes.
When the universe has reached equilibrium, the curvature perturbation $\zeta$ is equal to the perturbation of the logarithm of the scale factor $a$ on a constant energy density hypersurface,
$\zeta=\delta N\equiv \delta \ln a|_H,$
which one can find by solving the Friedmann equation for each universe separately.

At the end of inflation, all fields except light scalar fields vanish. It is therefore sufficient to consider the dependence of $N$ on the ``initial'' values of the light scalars $\phi_{\rm i}$ and $\chi_{\rm i}$ by which we mean their average values in the given Hubble volume at the end of inflation. The model is symmetric under $\chi_{\rm i}\rightarrow-\chi_{\rm i}$, which implies $\partial N/\partial \chi_{\rm i}=0$. We can therefore Taylor expand
\begin{eqnarray}
\label{equ:zetadef}
 \zeta&=&\frac{\partial N}{\partial \phi_{\rm i}}\delta\phi_{\rm i}
+\frac{1}{2}\frac{\partial^2 N}{\partial\phi_{\rm i}^2}
	\left(\delta\phi_{\rm i}^2-\langle\delta\phi_{\rm i}^2\rangle\right)
	\nonumber\\&&
+\frac{1}{2}\frac{\partial^2 N}{\partial\chi_{\rm i}^2}
	\left(\delta\chi_{\rm i}^2-\langle\delta\chi_{\rm i}^2\rangle\right)+O(\delta\phi_{\rm i}^3),
\end{eqnarray}
where we have subtracted the averages $\langle\delta\phi_{\rm i}^2\rangle$ and $\langle\delta\chi_{\rm i}^2\rangle$ to keep the unperturbed value of $\zeta$ zero.
Assuming, as we do, that the initial field values are Gaussian, the first term described the usual Gaussian contribution, and the second term is a local non-Gaussian contribution that is present already in single-field models. The third term, however, is more interesting, since it represents the contribution from the other light field $\chi$.
To measure the level of this contribution, we need to find 
the dependence of $N\equiv\ln a$ on the initial value of $\chi$.

Previous applications of the separate universe approach they have generally assumed that not only the metric but also the field configurations are homogeneous inside each separate universe~\cite{Tanaka:2003cka,Suyama:2006rk}. the dynamics are then described by a system of coupled ODEs,
\begin{eqnarray}
\label{equ:homogeqs}
\ddot\phi+3H\dot\phi+\frac{\partial V}{\partial \phi}&=&0,\nonumber\\
\ddot\chi+3H\dot\phi+\frac{\partial V}{\partial \chi}&=&0,\nonumber\\
H^2&=&\frac{\rho}{3M_{\rm Pl}^2},
\end{eqnarray}
where $\rho=\frac{1}{2}\dot\phi^2+\frac{1}{2}\dot\chi^2+V(\phi,\chi)$ is the energy density.
Solving these equations and calculating the scale factor $a$ at some suitably chosen $H$ gives a non-linear way of calculating the curvature perturbation $\zeta$.

The contribution by the $\chi$ field to the non-Gaussianity is not of the simple ``local'' type (\ref{equ:fNLorig}), but defining $f_{\rm NL}$ as a suitable ratio of the three-point and two-point correlation functions of $\zeta$, Boubekeur and Lyth obtained for this type of non-Gaussianity an effective value of~\cite{Boubekeur:2005fj}
\begin{eqnarray}
\label{equ:fNLcalc}
 f_{\rm NL}&\approx&-\frac{5}{48}\left(\frac{\partial^2N}{\partial\chi_{\rm i}^2}\right)^3
 \frac{{\cal P}_\chi^3}{{\cal P}_\zeta^2}
		\ln\frac{k}{H}
\nonumber\\&&
=-\frac{5}{9\pi^2}\left(\frac{\partial^2N}{\partial\chi_{\rm i}^2}\right)^3\lambda M_{\rm Pl}^6\ln\frac{k}{H},
\end{eqnarray}
where we have used the expressions 
${\cal P}_\chi=(H^2/4\pi^2)$
and ${\cal P}_\zeta=(V/24\pi^2\epsilon M_{\rm Pl}^4)$ for the power spectra of $\chi$ and $\zeta$.
The logarithm reflects an infrared divergence, which is cut off by the length scale at which the averages in (6) are computed, i.e., the maximum observable scale.
This expression is only valid if the non-Gaussianity is small, $f_{\rm NL}\lesssim 10^5$.
In any case, the observational limits on $f_{\rm NL}$ lead to the constraint
\begin{equation}
\label{equ:Nchiconst}
\left|\frac{\partial^2N}{\partial\chi_{\rm i}^2}\right|\lesssim 10^5.
\end{equation}

The key finding in earlier work on this model~\cite{Tanaka:2003cka,Suyama:2006rk} was that the dependence of $N$ on the initial $\chi$ appears random, suggesting chaotic dynamics. This would mean that the the contribution to the curvature perturbation is simply white noise, with no observable effects on large scales. As we will see, this chaotic behaviour is, in fact, merely a consequence of ignoring the inhomogeneous modes completely.
When they are included, individual degrees of freedom may still behave chaotically, but the averaged dynamics, which determines the expansion of the universe and thereby the curvature perturbation, is non-chaotic.

To include the effects of sub-Hubble inhomogeneities, we employ a hierarchy between 
the two relevant length scales. The length scales relevant for the non-equilibrium dynamics, which we call {\em microscopic}, are smaller than the horizon size, but we are interested in the curvature perturbation on {\em macroscopic}, super-Hubble scales that are observable today.

We approximate the microscopic dynamics by defining the inflaton and the other relevant fields as classical fields on a comoving lattice. The lattice size is chosen to be large enough to describe the microscopic dynamics accurately but smaller than the horizon at all times, so that the space-time geometry inside the lattice can be well approximated by the FRW metric. 
The simulations in Ref.~\cite{BasteroGil:2007mm} confirm that sub-horizon metric perturbations can be safely ignored.
We can therefore use the standard classical field theory methods that have been used to study preheating over the last decade.

Since the lattice size is larger than the characteristic length scale of the microscopic dynamics, the average energy density is the whole Hubble volume is well approximated by the average energy density $\overline{\rho}$ in our lattice,
given by
\begin{equation}
\overline\rho=\frac{1}{L^3}\!\int\! d^3x\left(\frac{\dot\phi^2}{2}+\frac{\dot\chi^2}{2}
+\frac{(\vec\nabla\phi)^2}{2a^2}+\frac{(\vec\nabla\chi)^2}{2a^2}
+V(\phi,\chi)\right),
\end{equation}
where the integration is over the lattice and $L$ is the co-moving lattice size.

We can now use $\overline\rho$ in the Friedmann equation to calculate the evolution of the scale factor. 
In a sense, we are therefore using the field theory simulation to calculate the equation of state inside the Hubble volume. 
This same approach has been used in the past to study the formation of primordial black holes during preheating~\cite{Suyama:2004mz}. 
The field evolution is given by the full classical equation of motion.
\begin{eqnarray}
\ddot\phi+3H\dot\phi-\frac{1}{a^2}\vec\nabla^2\phi+\frac{\partial V}{\partial \phi}&=&0,\nonumber\\
\ddot\chi+3H\dot\chi-\frac{1}{a^2}\vec\nabla^2\chi+\frac{\partial V}{\partial \chi}&=&0,\nonumber\\
H^2&=&\frac{\overline\rho}{3M_{\rm Pl}^2}.
\end{eqnarray}
We solved this system of equations in conformal time ($d\eta=a^{-1}dt$) using the second-order Runge-Kutta algorithm for the field equations coupled to an Euler method for the Friedmann equation~\cite{inprep}.

To set up the initial conditions, we followed the standard approach in studies of preheating~\cite{Khlebnikov:1996mc,Prokopec:1996rr}. We drew the initial values of the fields and their time derivatives from a Gaussian ensemble that has the same two-point functions as the quantum mechanical vacuum state,
\begin{eqnarray}
\label{equ:quantumfluct}
\langle \chi_k\chi_q\rangle&=&(2\pi)^3\delta(k+q)\frac{1}{2\omega_k},\nonumber\\
\langle \dot\chi_k\dot\chi_q\rangle&=&(2\pi)^3\delta(k+q)\frac{\omega_k}{2},
\end{eqnarray}
with all other two-point correlators vanishing and the $\phi$-field homogeneous at the beginning of the simulation. In the linearised classical theory, these two-point functions evolve exactly the same way as in the quantum theory. On the other hand, when the dynamics become non-linear, occupation numbers of the modes are generally large, and the classical field equations should again describe the dynamics very well. For each set of parameters, we repeated the simulation between 60 and 240 times with different random realisation of the initial fluctuations.


We fixed the inflaton self-coupling to $\lambda=7\times 10^{-14}$, and started the simulation at $\phi=5$ in Planck units. We used lattices of up to $32^3$ points with spacing $\delta x=1.25\times 10^5$ and time step $\delta t=4\times 10^2$.
For these parameters, the contribution to the energy density by the ``quantum'' fluctuations~(\ref{equ:quantumfluct}),
$\rho_{\rm fluct}\sim\delta x^{-4}\sim 10^{-21}$, is well below the physical energy density
$\rho_{\rm phys}\approx \lambda \phi_{\rm ini}^2\sim 10^{-11},$ and should not 
affect the dynamics.

\begin{figure}
	\vspace*{-.7cm}
 \includegraphics[width=8cm]{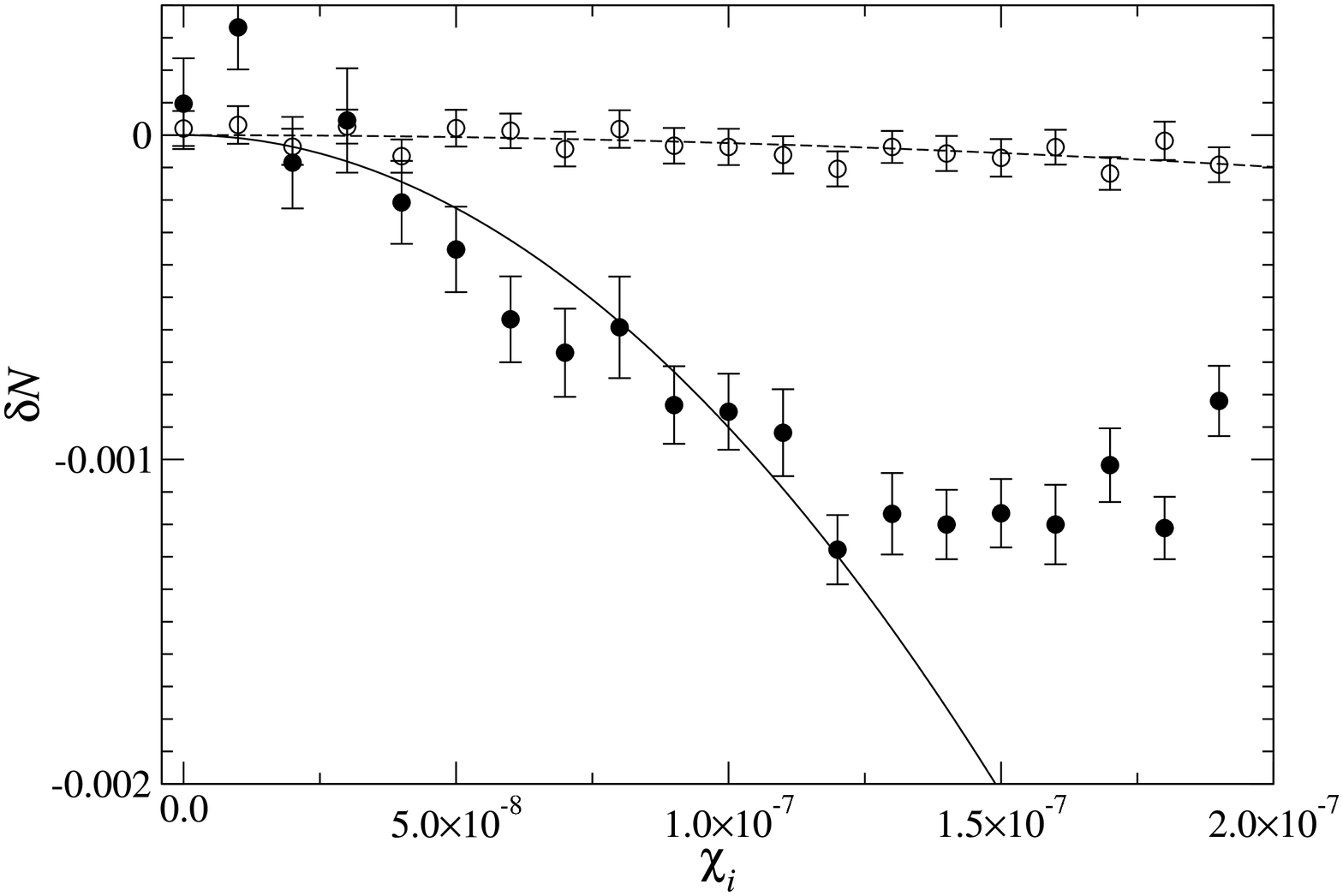}
\caption{\label{fig:Nvschi}The dependence of $N=\ln a$ on $\chi_{\rm i}$ measured at $H=5.53\times 10^{-12} M_{\rm Pl}$ for $g^2/\lambda=1.192$ (unfilled circles) and $g^2/\lambda=1.875$ (filled circles). The curves are quadratic fits (\ref{equ:quadfit}).}
\end{figure}

To determine the curvature perturbation $\zeta$ we have to measure the dependence of the number of e-foldings $N=\ln a$ on the initial value of $\chi$ at a suitable fixed value of the Hubble rate such that the system has reached a quasi-equilibrium state. Figure~\ref{fig:Nvschi} shows $N(\chi)$ measured at $H=5.53\times 10^{-12} M_{\rm Pl}$ for $g^2/\lambda=1.192$ and $g^2/\lambda=1.875$. The data appear smooth, within errors, in contrast with the previous separate universe calculations~\cite{Tanaka:2003cka,Suyama:2006rk}. There is significant scatter between individual runs, but since that variation is random and uncorrelated, it will contribute a white noise signal that is unobservable on large scales.

According to Eq.~(\ref{equ:zetadef}),
the new contribution to the curvature perturbation depends on the second derivative 
$\partial^2 N/\partial\chi_{\rm i}^2$ where $\chi_{\rm i}$ is the value of the zero mode at the beginning of the simulation. We measure it by fitting the data at low $\chi_{\rm i}$ with a quadratic function,
\begin{equation}
\label{equ:quadfit}
 N(\chi_{\rm i})=N(0)+c\chi_{\rm i}^2,
\end{equation}
so that $\partial^2N/\partial\chi_{\rm i}^2=2c$.
The curves in Fig.~\ref{fig:Nvschi} show the fits for $g^2/\lambda=1.192$ and $g^2/\lambda=1.875$.

\begin{table}
 \begin{tabular}{c|c|c}
 $g^2/\lambda$ & $\partial^2N/\partial\chi_{\rm i}^2$ & $\mu(\kappa=0)$
 \cr
 \hline
 $1.05$ & $-10^{4.60\pm0.05}$ & $0.085$\cr
 $1.192$ & $10^{9.69\pm0.12}$ & $0.157$\cr
 $1.875$ & $10^{11.26\pm0.05}$ & $0.237$\cr
 $2.7$ & $10^{14.01\pm0.10}$ & $0.157$
 \end{tabular}
\caption{\label{fig:cvsg}The second derivative $\partial^2N/\partial\chi_{\rm i}^2$ determined from the fit (\ref{equ:quadfit}) at $H=5.53\times 10^{-12}M_{\rm Pl}$ for different values of $g^2/\lambda$. The third column shows the Floquet index of the homogeneous mode.}
\end{table}

Table~\ref{fig:cvsg} shows the fitted values of $\partial^2 N/\partial\chi_{\rm i}^2$ for four different couplings $g^2/\lambda$. These values do not depend significantly on the value of $H$ at which they are measured. At $g^2/\lambda=1.05$ the measured value corresponds to $f_{\rm NL}\sim O\left( 1 \right)$. This is within observational bounds~(\ref{equ:Nchiconst}) and would be observable in future experiments, whereas the three other points are excluded by current observations. In fact, they are so large that the contribution from preheating dominates the curvature perturbation. Even the spectrum would therefore disagree with observations. 

Comparison of the three largest values of $g^2/\lambda$ in Table~\ref{fig:cvsg} demonstrates the importance of the inhomogeneous modes for the effect. The resonance is at its strongest at $g^2/\lambda=1.875$, and earlier studies~\cite{Tanaka:2003cka,Suyama:2006rk,Jokinen:2005by} which
ignored inhomogeneous modes would have suggested that effect would be strongest at this parameter value. Likewise, they would have suggested that the parameter values $g^2/\lambda=2.7$ and $g^2/\lambda=1.192$ would have produced a smaller effect because the resonance is weaker.
However, our results show strongest effect at $g^2/\lambda=2.7$ and weaker at $g^2/\lambda=1.192$. 

This finding has a simple explanation. What matters is not the absolute strength of the resonance, but the relative growth rates of the zero mode and the inhomogeneous modes. Near the lower edge of the resonance band $g^2/\lambda=1$, the growth rate of the zero mode is very small but inhomogeneous modes grow with a significant rate. Hence, $\partial^2N/\partial\chi_{\rm i}^2$ is very small. As $g^2/\lambda$ is increased, the growth rate of the zero mode grows, and $\partial^2N/\partial\chi_{\rm i}^2$ grows very rapidly. At $g^2/\lambda\approx1.4$, the two growth rates are equal, and the dependence on $g^2/\lambda$ is weaker. Above this value, $\partial^2N/\partial\chi_{\rm i}^2$ starts to grow more rapidly again, as the growth rate of the inhomogeneous modes starts to fall.
This agrees with our findings, and it will be very interesting to map this behaviour out including the sign of $\partial^2N/\partial\chi_{\rm i}^2$ in detail in future studies \cite{inprep}.

To summarise, we have demonstrated a method for computing  the curvature perturbation non-linearly. The method can be readily applied to any inflationary model in which the dynamics are dominated by bosonic fields. It requires an average over random initial conditions and therefore the results will inevitably have statistical errors, but as our results show, it is possible to achieve reasonably good signal-to-noise ratios with existing computing technology.

The strength of the mechanism that generates the curvature perturbations during preheating depends on how sensitive the expansion of the universe is on the initial value of the homogeneous mode. This is determined by the relative growth rates of the zero mode and the inhomogeneous modes. Thus, calculations that ignore the effect of inhomogeneous modes cannot be trusted.

Our results rule out parts of the parameter space in the massless preheating model, because they would lead to extremely high level of non-Gaussian curvature perturbations, but they also show that
others are compatible with observations. They demonstrate that reheating mechanisms can be crucial for the viability of inflationary models.

This work was supported by STFC and made use of the Imperial College High Performance Computing facilities.

\bibliography{paper}

\begin{thebibliography}{21}
\expandafter\ifx\csname natexlab\endcsname\relax\def\natexlab#1{#1}\fi
\expandafter\ifx\csname bibnamefont\endcsname\relax
  \def\bibnamefont#1{#1}\fi
\expandafter\ifx\csname bibfnamefont\endcsname\relax
  \def\bibfnamefont#1{#1}\fi
\expandafter\ifx\csname citenamefont\endcsname\relax
  \def\citenamefont#1{#1}\fi
\expandafter\ifx\csname url\endcsname\relax
  \def\url#1{\texttt{#1}}\fi
\expandafter\ifx\csname urlprefix\endcsname\relax\def\urlprefix{URL }\fi
\providecommand{\bibinfo}[2]{#2}
\providecommand{\eprint}[2][]{\url{#2}}

\bibitem[{\citenamefont{Komatsu and Spergel}(2000)}]{Komatsu:2000vy}
\bibinfo{author}{\bibfnamefont{E.}~\bibnamefont{Komatsu}} \bibnamefont{and}
  \bibinfo{author}{\bibfnamefont{D.~N.} \bibnamefont{Spergel}}
  (\bibinfo{year}{2000}), \eprint{astro-ph/0012197}.

\bibitem[{\citenamefont{Maldacena}(2003)}]{Maldacena:2002vr}
\bibinfo{author}{\bibfnamefont{J.~M.} \bibnamefont{Maldacena}},
  \bibinfo{journal}{JHEP} \textbf{\bibinfo{volume}{05}}, \bibinfo{pages}{013}
  (\bibinfo{year}{2003}).

\bibitem[{\citenamefont{Boubekeur and Lyth}(2006)}]{Boubekeur:2005fj}
\bibinfo{author}{\bibfnamefont{L.}~\bibnamefont{Boubekeur}} \bibnamefont{and}
  \bibinfo{author}{\bibfnamefont{D.~H.} \bibnamefont{Lyth}},
  \bibinfo{journal}{Phys. Rev.} \textbf{\bibinfo{volume}{D73}},
  \bibinfo{pages}{021301} (\bibinfo{year}{2006}).

\bibitem[{\citenamefont{Creminelli et~al.}(2007)\citenamefont{Creminelli,
  Senatore, Zaldarriaga, and Tegmark}}]{Creminelli:2006rz}
\bibinfo{author}{\bibfnamefont{P.}~\bibnamefont{Creminelli}},
  \bibinfo{author}{\bibfnamefont{L.}~\bibnamefont{Senatore}},
  \bibinfo{author}{\bibfnamefont{M.}~\bibnamefont{Zaldarriaga}},
  \bibnamefont{and} \bibinfo{author}{\bibfnamefont{M.}~\bibnamefont{Tegmark}},
  \bibinfo{journal}{JCAP} \textbf{\bibinfo{volume}{0703}}, \bibinfo{pages}{005}
  (\bibinfo{year}{2007}).

\bibitem[{\citenamefont{Bassett et~al.}(2006)\citenamefont{Bassett, Tsujikawa,
  and Wands}}]{Bassett:2005xm}
\bibinfo{author}{\bibfnamefont{B.~A.} \bibnamefont{Bassett}},
  \bibinfo{author}{\bibfnamefont{S.}~\bibnamefont{Tsujikawa}},
  \bibnamefont{and} \bibinfo{author}{\bibfnamefont{D.}~\bibnamefont{Wands}},
  \bibinfo{journal}{Rev. Mod. Phys.} \textbf{\bibinfo{volume}{78}},
  \bibinfo{pages}{537} (\bibinfo{year}{2006}).

\bibitem[{\citenamefont{Prokopec and Roos}(1997)}]{Prokopec:1996rr}
\bibinfo{author}{\bibfnamefont{T.}~\bibnamefont{Prokopec}} \bibnamefont{and}
  \bibinfo{author}{\bibfnamefont{T.~G.} \bibnamefont{Roos}},
  \bibinfo{journal}{Phys. Rev.} \textbf{\bibinfo{volume}{D55}},
  \bibinfo{pages}{3768} (\bibinfo{year}{1997}).

\bibitem[{\citenamefont{Greene et~al.}(1997)\citenamefont{Greene, Kofman,
  Linde, and Starobinsky}}]{Greene:1997fu}
\bibinfo{author}{\bibfnamefont{P.~B.} \bibnamefont{Greene}},
  \bibinfo{author}{\bibfnamefont{L.}~\bibnamefont{Kofman}},
  \bibinfo{author}{\bibfnamefont{A.~D.} \bibnamefont{Linde}}, \bibnamefont{and}
  \bibinfo{author}{\bibfnamefont{A.~A.} \bibnamefont{Starobinsky}},
  \bibinfo{journal}{Phys. Rev.} \textbf{\bibinfo{volume}{D56}},
  \bibinfo{pages}{6175} (\bibinfo{year}{1997}).

\bibitem[{\citenamefont{Bassett and Viniegra}(2000)}]{Bassett:1999cg}
\bibinfo{author}{\bibfnamefont{B.~A.} \bibnamefont{Bassett}} \bibnamefont{and}
  \bibinfo{author}{\bibfnamefont{F.}~\bibnamefont{Viniegra}},
  \bibinfo{journal}{Phys. Rev.} \textbf{\bibinfo{volume}{D62}},
  \bibinfo{pages}{043507} (\bibinfo{year}{2000}).

\bibitem[{\citenamefont{Enqvist et~al.}(2005)\citenamefont{Enqvist, Jokinen,
  Mazumdar, Multamaki, and Vaihkonen}}]{Enqvist:2004ey}
\bibinfo{author}{\bibfnamefont{K.}~\bibnamefont{Enqvist}},
  \bibinfo{author}{\bibfnamefont{A.}~\bibnamefont{Jokinen}},
  \bibinfo{author}{\bibfnamefont{A.}~\bibnamefont{Mazumdar}},
  \bibinfo{author}{\bibfnamefont{T.}~\bibnamefont{Multamaki}},
  \bibnamefont{and}
  \bibinfo{author}{\bibfnamefont{A.}~\bibnamefont{Vaihkonen}},
  \bibinfo{journal}{Phys. Rev. Lett.} \textbf{\bibinfo{volume}{94}},
  \bibinfo{pages}{161301} (\bibinfo{year}{2005}).

\bibitem[{\citenamefont{Jokinen and Mazumdar}(2006)}]{Jokinen:2005by}
\bibinfo{author}{\bibfnamefont{A.}~\bibnamefont{Jokinen}} \bibnamefont{and}
  \bibinfo{author}{\bibfnamefont{A.}~\bibnamefont{Mazumdar}},
  \bibinfo{journal}{JCAP} \textbf{\bibinfo{volume}{0604}}, \bibinfo{pages}{003}
  (\bibinfo{year}{2006}).

\bibitem[{\citenamefont{Erd\'{e}lyi}(1953)}]{htbook}
\bibinfo{author}{\bibfnamefont{A.}~\bibnamefont{Erd\'{e}lyi}},
  \emph{\bibinfo{title}{Higher Transcendental Functions, Volume 2}}
  (\bibinfo{publisher}{McGraw-Hill, New York}, \bibinfo{year}{1953}).

\bibitem[{\citenamefont{Traschen and Brandenberger}(1990)}]{Traschen:1990sw}
\bibinfo{author}{\bibfnamefont{J.~H.} \bibnamefont{Traschen}} \bibnamefont{and}
  \bibinfo{author}{\bibfnamefont{R.~H.} \bibnamefont{Brandenberger}},
  \bibinfo{journal}{Phys. Rev.} \textbf{\bibinfo{volume}{D42}},
  \bibinfo{pages}{2491} (\bibinfo{year}{1990}).

\bibitem[{\citenamefont{Kofman et~al.}(1994)\citenamefont{Kofman, Linde, and
  Starobinsky}}]{Kofman:1994rk}
\bibinfo{author}{\bibfnamefont{L.}~\bibnamefont{Kofman}},
  \bibinfo{author}{\bibfnamefont{A.~D.} \bibnamefont{Linde}}, \bibnamefont{and}
  \bibinfo{author}{\bibfnamefont{A.~A.} \bibnamefont{Starobinsky}},
  \bibinfo{journal}{Phys. Rev. Lett.} \textbf{\bibinfo{volume}{73}},
  \bibinfo{pages}{3195} (\bibinfo{year}{1994}).

\bibitem[{\citenamefont{Felder and Navros}(2007)}]{Felder:2007ef}
\bibinfo{author}{\bibfnamefont{G.~N.} \bibnamefont{Felder}} \bibnamefont{and}
  \bibinfo{author}{\bibfnamefont{O.}~\bibnamefont{Navros}},
  \bibinfo{journal}{JCAP} \textbf{\bibinfo{volume}{0702}}, \bibinfo{pages}{014}
  (\bibinfo{year}{2007}).

\bibitem[{\citenamefont{Bastero-Gil et~al.}(2007)\citenamefont{Bastero-Gil,
  Tristram, Macias-Perez, and Santos}}]{BasteroGil:2007mm}
\bibinfo{author}{\bibfnamefont{M.}~\bibnamefont{Bastero-Gil}},
  \bibinfo{author}{\bibfnamefont{M.}~\bibnamefont{Tristram}},
  \bibinfo{author}{\bibfnamefont{J.~F.} \bibnamefont{Macias-Perez}},
  \bibnamefont{and} \bibinfo{author}{\bibfnamefont{D.}~\bibnamefont{Santos}}
  (\bibinfo{year}{2007}), \eprint{arXiv:0709.3510 [astro-ph]}.

\bibitem[{\citenamefont{Salopek and Bond}(1990)}]{Salopek:1990jq}
\bibinfo{author}{\bibfnamefont{D.~S.} \bibnamefont{Salopek}} \bibnamefont{and}
  \bibinfo{author}{\bibfnamefont{J.~R.} \bibnamefont{Bond}},
  \bibinfo{journal}{Phys. Rev.} \textbf{\bibinfo{volume}{D42}},
  \bibinfo{pages}{3936} (\bibinfo{year}{1990}).

\bibitem[{\citenamefont{Tanaka and Bassett}(2003)}]{Tanaka:2003cka}
\bibinfo{author}{\bibfnamefont{T.}~\bibnamefont{Tanaka}} \bibnamefont{and}
  \bibinfo{author}{\bibfnamefont{B.}~\bibnamefont{Bassett}}
  (\bibinfo{year}{2003}), \eprint{astro-ph/0302544}.

\bibitem[{\citenamefont{Suyama and Yokoyama}(2007)}]{Suyama:2006rk}
\bibinfo{author}{\bibfnamefont{T.}~\bibnamefont{Suyama}} \bibnamefont{and}
  \bibinfo{author}{\bibfnamefont{S.}~\bibnamefont{Yokoyama}},
  \bibinfo{journal}{Class. Quant. Grav.} \textbf{\bibinfo{volume}{24}},
  \bibinfo{pages}{1615} (\bibinfo{year}{2007}).

\bibitem[{\citenamefont{Suyama et~al.}(2005)\citenamefont{Suyama, Tanaka,
  Bassett, and Kudoh}}]{Suyama:2004mz}
\bibinfo{author}{\bibfnamefont{T.}~\bibnamefont{Suyama}},
  \bibinfo{author}{\bibfnamefont{T.}~\bibnamefont{Tanaka}},
  \bibinfo{author}{\bibfnamefont{B.}~\bibnamefont{Bassett}}, \bibnamefont{and}
  \bibinfo{author}{\bibfnamefont{H.}~\bibnamefont{Kudoh}},
  \bibinfo{journal}{Phys. Rev.} \textbf{\bibinfo{volume}{D71}},
  \bibinfo{pages}{063507} (\bibinfo{year}{2005}).

\bibitem[{\citenamefont{Chambers and Rajantie}()}]{inprep}
\bibinfo{author}{\bibfnamefont{A.}~\bibnamefont{Chambers}} \bibnamefont{and}
  \bibinfo{author}{\bibfnamefont{A.}~\bibnamefont{Rajantie}},
  \emph{\bibinfo{title}{in preparation}}.

\bibitem[{\citenamefont{Khlebnikov and Tkachev}(1996)}]{Khlebnikov:1996mc}
\bibinfo{author}{\bibfnamefont{S.~Y.} \bibnamefont{Khlebnikov}}
  \bibnamefont{and} \bibinfo{author}{\bibfnamefont{I.~I.}
  \bibnamefont{Tkachev}}, \bibinfo{journal}{Phys. Rev. Lett.}
  \textbf{\bibinfo{volume}{77}}, \bibinfo{pages}{219} (\bibinfo{year}{1996}).

\end{thebibliography}

\end{document}